\documentclass[prb]{revtex4}
\begin{document}

\begin{large}

\begin{center}
\begin{Large}
{\Large\bf{Surface order-disorder phase transitions and
percolation}}
\end{Large}
\end{center}

\begin{center}
M.C. Gim\'enez, F. Nieto and A. J. Ramirez--Pastor$^{\dag}$
\end{center}

\begin{center} {\it Departamento de F\'{\i}sica,
Universidad Nacional de San Luis, CONICET, \\ Chacabuco 917,
D5700BWS, San Luis, Argentina \\ cecigime@unsl.edu.ar;
fnieto@unsl.edu.ar; antorami@unsl.edu.ar}
\end{center}

\vspace{0cm}
\begin{center} Abstract \end{center}

In the present paper, the connection between surface
order-disorder phase transitions and the percolating properties of
the adsorbed phase has been studied. For this purpose, four
lattice-gas models in presence of repulsive interactions have been
considered. Namely, monomers on honeycomb, square and triangular
lattices, and dimers (particles occupying two adjacent adsorption
sites) on square substrates. By using Monte Carlo simulation and
finite-size scaling analysis, we obtain the percolation threshold
$\theta_c$ of the adlayer, which presents an interesting
dependence with $w/k_BT$ (being $w$, $k_B$ and $T$, the lateral
interaction energy, the Boltzmann's constant and temperature,
respectively). For each geometry and adsorbate size, a phase
diagram separating a percolating and a non-percolating region is
determined.

\vspace{1cm}


\vspace{0.0cm}

\\  \rm

\vspace{0.0cm}

\noindent $\dag$ To whom all correspondence should be addressed.

\newpage

\section{Introduction}

Despite over three decades of intensive work, the interplay between
percolation properties and thermal phase transitions is still an
open problem. In this sense, the study of geometrical structures
close to the critical point allows a better understanding of the
mechanism of the phase transition
\cite{coniglio1,coniglio2,coniglio3}. The geometric critical
phenomena exemplified by percolation possess many striking parallels
with the thermally
 driven critical phenomena, as it is provided by the Fortuin-Kasteleyn mapping \cite{FK}.
Cluster description of thermodynamic phase transitions have been
used since long time to elucidate the nature of the transitions by
providing a geometrical interpretation of density correlations
\cite{Hill}. Fisher has introduced the phenomenological droplet
model in which the fluctuations are associated to clusters or
droplets which percolate at critical point \cite{Fisher}. More
recently, it was concluded that the percolation transition can be
considered as a particular case of the $q$-state
 Potts model where $q$ is equal to 1 \cite{FK01,Wu01}, and can be described as a second order one,
 whose universality class depends only upon the space dimensionality.
By following this line of reasoning, a wide variety of systems have
been studied. Among the more recent contributions, the behavior of
colloids and gels has been discussed in an interesting paper by A.
Coniglio \cite{coniglio1}. The author concluded that it is very
important to define the appropriate cluster for each phenomenon.
Later, an important contribution has been made by S. Fortunato
\cite{fortu1,fortu2} who analyzes the critical exponents of both the
thermal and percolation phase transitions occurring for different
models in two dimensions.

 Although most of the works on the
subject are devoted to the study of lattice-gas models in presence
of attractive lateral interactions (or ferromagnetic coupling in
magnetic language), there have been a few studies related to
repulsive interactions and order-disorder phase transitions. In the
last case, the definitions of connectivity and cluster of particles
belonging to the adsorbed phase are not the same and the
relationship between percolating clusters and critical points is
non-trivial. In a previous paper \cite{gimenez}, we studied the
percolation of monomers on a square lattice as the particles
interact with repulsive energies. The present contribution goes a
step further, including honeycomb and triangular substrates and
multisite-occupancy \cite{RUDEVE,SURFSCI3,PRB3,PRB4,PRB5}
(adsorbates occupying more than one site).

The outline of the paper is as follows: In Section \ref{model} we
describe the lattice-gas model and the finite-size scaling theory.
In Section \ref{pd} we present and discuss the results along with
general conclusions.

\section{Model and finite-size scaling theory}\label{model}

In order to consolidate the ideas involved here, four different
physical systems have been considered, according to the adsorbate's
size and surface geometry:

\noindent {\rm Model I): Monomers adsorbed on square lattices.}

\noindent {\rm Model II): Monomers adsorbed on honeycomb
lattices.}

\noindent {\rm Model III): Monomers adsorbed on triangular
lattices.}

\noindent {\rm Model IV): Dimers adsorbed on square lattices.}

In all cases, the substrate is represented by $M=L \times L$
equivalent adsorption sites with periodic boundary conditions.

To describe the system of $N$ particles adsorbed on $M$ sites at a
given temperature $T$, let us introduce the occupation variable
$c_i$ which can take the following values: $c_i=0$ if the
corresponding site is empty and $c_i = 1$ if the site is  occupied
by an adatom (or dimer unit). Under these considerations, the
Hamiltonian of the system is given by,
\begin{equation}
H = w \sum_{_{\langle i,j \rangle}} c_i c_j - N(k-1) w +
\epsilon_o \sum_{i}^M  c_i \label{h}
\end{equation}
where $w$ is the nearest-neighbor (NN) interaction energy (we
focus on the case of repulsive lateral interactions among adsorbed
particles, $w>0$); $\langle i,j \rangle$ represents pairs of $NN$
sites and $k=1(2)$ for monomers(dimers). The term $N(k-1)w$ is
subtracted in eq. (\ref{h}) since, in the case of $k=2$, the
summation over all the pairs of NN sites overestimates the total
energy by including $N$ bonds belonging to the $N$ adsorbed
dimers. Finally, $\epsilon_o $ is the  adsorption energy of the
sites on the surface (we have taken $\epsilon_o=0$ without loss of
generality).

For fixed values of surface coverage, $\theta = kN/M$, and
temperature $T$, the thermodynamic equilibrium is reached in the
canonical ensemble by using a standard Kawasaki algorithm
\cite{kawasaki,gimenez}. Thus, a set of $m=2000$ samples in thermal
equilibrium is generated by taking configurations separated from
each other by $1000$ Monte Carlo steps in order to avoid memory
effects.

The central idea of the percolation theory is based on finding the
minimum concentration $\theta$ for which at least a cluster [a group
of occupied sites in such a way that each site has at least one
occupied nearest-neighbor site] extends from one side to the
opposite one of the system. This particular value of the
concentration rate is named {\it critical concentration} or {\it
percolation threshold} and determines a phase transition in the
system.

It is well known that it is a quite difficult matter to analytically
determine the value of the percolation threshold for a given lattice
\cite{Stauffer,Sahimi,Zallen,Essam,Hovi}. Thus, in most cases,
percolation thresholds have to be estimated numerically by means of
computer simulations.

As the scaling theory predicts \cite{Binder}, the larger the system
size to study, the more accurate the values of the threshold
obtained therefrom. Thus, the finite-size scaling theory give us the
basis to achieve the percolation threshold and the critical
exponents of a system with a reasonable accuracy. For this purpose,
the probability $R=R^X _L(\theta)$ that a lattice composed of $L
\times L$ elements (sites or bonds) percolates at concentration
$\theta$ can be defined \cite{Stauffer}. Here,  the following
definitions can be given according to the meaning of $X$: a) $R^{R
(D)} _L(\theta)=$ the probability of finding a rightward (downward)
percolating cluster; b) $R^{I} _L(\theta)=$ the probability that we
find a cluster which percolates both in a rightward {\bf and} in a
downward direction; c) $R^{U} _L(\theta)=$ the probability of
finding either a rightward {\bf or} a downward percolating cluster
and d) $R^{A} _L(\theta) \equiv \frac{1}{2} \left[R^{R}
_L(\theta)+R^{D} _L(\theta) \right] \equiv \frac{1}{2} \left[R^{I}
_L(\theta)+ R^{U} _L(\theta) \right] $. Based on these definitions
and using the methodology described in
Refs.~[\onlinecite{gimenez,YONE1,YONE2,COR1,COR2,DOLZ,DOLZ01,Watts,Lag01,Lag02,Cardy}],
the percolation thresholds were calculated by extensive use of
finite size scaling techniques. The interested reader is urged to
read the above cited articles for a more complete discussion of this
issue.

\section{Percolation phase diagram: Results and conclusions}\label{pd}

By using the scheme discussed above, the critical curves, $\theta_c$
vs. $K$ (being $K \equiv w/k_BT$), separating the percolating and
non-percolating regions, were calculated.

In Fig.~1, the percolation phase diagram is shown for Model I. As
$K=0$ (non-interacting adsorbate), the adsorption-desorption process
reproduces a random deposition, which is fully equivalent to random
percolation. Consequently, the expected value of $\theta_c=0.592$ is
reached for $K =0$. As $K$ is increased, two well-differentiated
regimes can be distinguished: $i)$ from $K = 0$ up to $K \approx
1.76$ (being $K \approx 1.76$ the reduced critical temperature for
the order-disorder phase transition occurring in the system),
$\theta_c$ increases linearly with $K$; and $ii)$ for $K > 1.76$,
$\theta_c$ remains constant as $K$ is increased. This behavior can
be explained from simple geometrical arguments. Namely, lateral
repulsive couplings avoid the occupation of nearest-neighbors sites,
and consequently, increase the percolation threshold. In the limit
case, once $K_c$ is reached, the adlayer does not vary significantly
as $K$ is increased, and $\theta_c$ reaches its saturation value,
being $\theta_c  \approx 0.66$ for $K
> 1.76$. It is worth to emphasize that the presence of strong lateral
interactions (and consequently, the existence of a phase transition
occurring in the system) yields an increase in the computational
effort to get accuracy values of the percolation threshold.

It is important to bear in mind that the points in Fig.~1 correspond
to states in thermal equilibrium. In order to reflect this
situation, we have calculated the adsorption isotherms (mean
coverage as a function of the reduced chemical potential,
$\mu/k_BT$) for repulsively interacting adparticles in a wide range
of temperatures. The adsorption process was simulated through a
Grand Canonical Ensemble Monte Carlo (GCEMC) method. Relaxation
toward equilibrium relied upon Glauber dynamics \cite{Nicholson}.

For a given value of temperature $T$ and chemical potential $\mu$,
an initial configuration with $N$ monomers (dimers) adsorbed at
random positions on $N$ ($2N$) sites is generated. Then an
adsorption-desorption process is started, where a site (pair of
nearest-neighbor sites) is chosen at random and an attempt is made
to change its occupancy state with probability given by the
Metropolis rule \cite{Metropolis}:
\begin{equation}
P = \min \left\{1,\exp\left( -\Delta \tilde{H} / k_B T\right)
\right\}
\end{equation}
where $\tilde{H}=\tilde{H}_f -\tilde{H}_i$ is the difference between
the effective Hamiltonians of the final and initial states, being
$\tilde{H}=H-\mu \sum c_i$. A Monte Carlo Step (MCS) is achieved
when $N$ sites (pair of sites) have been tested to change its
occupancy state. The equilibrium state can be well reproduced after
discarding the first $m'=10^5-10^6$ MCS. Then, averages are taken
over $m=10^5-10^6$ successive configurations. In this framework, the
mean coverage is obtained as:
\begin{equation}
\theta = \frac{1}{N} \sum_i^N <c_i>
\end{equation}
where the thermal average $\langle ... \rangle$, means the time
average over the Monte Carlo simulation run.

In order to compare our numerical results with a theoretical
prediction, we have used one of the most reliable methods for
studying the thermodynamic properties of a system suffering a phase
transition: the Real Space Renormalization Group (RSRG)
\cite{Wilson}. The interested reader is referred to
Ref.~[\onlinecite{Wilson,Nie75}] for a detailed description of the
RSRG method and to
Refs.~[\onlinecite{Mahan,BerkerI,BerkerII,BerkerIII}] for
applications of the RSRG method to lattice gas models.

In the RSRG method developed by Niemeyer and van Leeuwen
\cite{Nie74} and Nauenberg and Nienhuis \cite{Nau74,Nau75}, the
whole lattice is divided into blocks (or cells) of $L$ sites. A
block spin $S_{\alpha}$ is assigned to each block. All blocks
together must form a square lattice with the lattice constant
$\sqrt{L}a$. The RSRG transformation of the spin system allows the
reduction of the number of independent variables, i.e. the
transition from the set of $N$ site spins $\{s_i\}$ to $N/L$ block
spins $\{S_{\alpha }\}$.We note that two values of the block spin
$S_{\alpha }=\pm 1$ corresponds to $2^L$ site spin configurations
(since $L$ spins are combined to form a block).  For blocks with odd
number of spins $S_{\alpha }$ is usually determined by the so-called
``majority rule'' (MR) \cite{Nie75}. For even  $L$ a rule must be
introduced in order to assign a definite value of the block spin to
any given configuration with the sum of site spins equal to zero. In
any case an obvious condition must be fulfilled: if the site spin
configuration $\{s_1,s_2\ldots s_L\}$ is assigned to a block spin
$S_{\alpha}$ with weighting factor $P$, then the configuration
$\{-s_1,-s_2\ldots -s_L\}$ is assigned to the $-S_{\alpha }$ domain
with the same $P$.

In the framework of the RSRG approach, one usually employs periodic
boundary conditions. It is assumed that the whole lattice is given
by the periodic continuation of a small cluster of blocks. In our
calculations we consider the smallest possible cluster of two
blocks. Due to the simplicity of this cluster, no additional
interactions appear in the renormalized Hamiltonian. It is the same
Hamiltonian of the square Ising spin system with, however,
renormalized values for the external magnetic field and for the pair
interaction parameter.

As was shown by Nauenberg and Nienhuis \cite{Nau74}, the free energy
of the system for any values of magnetic field and interaction
parameter can be evaluated in a series of sequential RSRG
transformations of the original Hamiltonian. The interested reader
will find details of the application of this technique for Models I,
II and III in the following references: square lattice,
Ref.~[\onlinecite{sl}], triangular lattice, Ref.~[\onlinecite{tl}]
and honeycomb lattice Ref.~[\onlinecite{hl}].


The results, obtained by Real Space Renormalization Group, RSRG
(solid lines) and Monte Carlo, MC, methods (small squares), are
shown in Fig.~2 \cite{sl}. At high temperatures the isotherms are
close to the Langmuir case (lattice-gas without lateral
interaction), i.e.
\begin{equation}
\theta (\mu)=\frac{\exp \beta(\mu+\varepsilon_o)}{1+\exp \beta
(\mu+\varepsilon_o)}.
\end{equation}
At low temperatures a broad plateau occurs around half coverage.
This plateau corresponds to the c$(2\times 2)$ ordered lattice-gas
phase (or, in magnetic language, to the AF ordered two-dimensional
spin system). Large spheres in Fig.~2 are the same points as in
Fig.~1. It is clear that such line avoids to enter in the region of
the coexistence of phases.

To reinforce the above result, in Fig.~3 the percolation line is
plotted (spheres) together with the coexistence curve in the
temperature-concentration diagram, which limits the region where the
$c(2 \times 2)$ ordered phase percolates. As it can be observed, the
percolation line remains in the region where the system is
disordered. These features clearly reveal that the definitions of
connectivity (in the sense of standard random percolation) and
therefore the definitions of the clusters of $c(2\times 2)$ ordered
structure are not the same.

In the case of Model II (Fig.~4), the general trend is similar to
that of the square lattice. The curve grows monotonically up to a
value of $K \approx 2$, where it reaches an almost constant value of
$0.75$ for the critical coverage degree. The explanation of this
trends is similar to the first case. The percolation line (spheres)
plotted together with the adsorption isotherms for repulsively
interacting particles adsorbed on a honeycomb lattice is shown in
Fig.~5. In Fig.~6 the same line is plotted in conjunction with the
phase diagram. Again, the percolation line remains in the region
where the system is disordered.

For Model III (Fig.~7),  it can be observed that the value of the
percolation threshold is near $0.5$ in the whole range of $K$
(notice the scale in the graph). A complete understanding of the
phase diagram is a very important help in the description of the
peculiarities of the temperature dependence of the percolation
threshold. In order to explain the antiferromagnetic ordering we
recall that a triangular lattice can be seen as a system composed of
three equivalent triangular sublattices. As is well known, pairwise
interaction results in a symmetrical phase diagram around $\theta =
0.5 $. For this lattice-gas system, triangular antiferromagnetic
lattice-gas, the phase diagram ($T,\theta$) consists of two
symmetrical curves around $\theta=0.5$. For $\theta \le 1/3$, the
ordered phase reveals that the particles are arranged in such a way
that pairs of particles on nearest-neighbor lattice sites are not
present.  In fact, most of the adsorbed particles are located in
only one of the three equivalent triangular sublattices, thus
avoiding possible interactions with other particles (which is
equivalent to an ordered $\uparrow\downarrow\downarrow $ phase using
magnetic language). This ordered phase prevails over the range
$\theta\le 0.5$ as is indicated in Fig.~8.

The symmetric branch of the phase diagram reflects the ordered phase
where two of the sublattices are occupied
($\uparrow\uparrow\downarrow $ in the magnetic language). This phase
diagram has been investigated by Schick, Walker and Wortis
\cite{Sch76,Sch77} by using RSRG as well as by the transfer matrix
method \cite{Kin81} and MC simulations \cite{tl,Met73}.

It is interesting to note that for $\theta\le 0.5$ a) the
corresponding ordered phase percolates  but b) there is not standard
percolation of the adsorbed monomers on the lattice. The percolation
line dividing the percolating and non percolating area yields in the
disordered phase of the antiferromagnetic phase diagram (see
Fig.~8). Furthermore, in the whole range of temperature, the
percolation occurs for a coverage lower than what is needed to built
the ordered phase for $\theta \ge 0.5$

Fig.~9 shows the percolation line together with the adsorption
isotherms for repulsively interacting particles. The adsorption
isoterms present clearly defined plateaus located at
$\theta=\frac13$ and $\theta=\frac23$. Strong enough repulsion
produces ordered phases when particles occupy preferentially sites
of a single sublattice ($\theta=\frac13$), or two sublattices
($\theta=\frac23$).

In the case of dimers on square lattices (Fig.~10), the curve
$\theta_c$ vs. $K$ (spheres) is similar to that of Fig.~1 (open
circles). Thus, $\theta_c$ grows linearly from $\theta_c(K=0)=0.56$
(as it is expected for random percolation of dimers) to a saturation
value close to $2/3$. This behavior has interesting consequences on
the temperature-concentration phase diagram. In fact, as it has been
reported in the literature \cite{SURFSCI3}, a ``zig-zag" (ZZ)
ordered phase, characterized by domains of parallel ZZ strips
oriented at $\pm 45^o$ from the lattice symmetry axes, separated
from each other by strips of single empty sites, was found at $2/3$
monolayer coverage (Fig.~11). The ordered phase is separated from
the disordered state by a order-disorder phase transition occurring
at a finite critical temperature. An accurate determination of this
critical temperature has been recently obtained
[$k_BT_c/w=0.182(1)$] \cite{EURLET}.

A simple inspection of Fig.~11 shows the existence of long-range
connectivity for the low-temperature phase at $2/3$ coverage. This
finding, along with the tendency to $2/3$ of the curve in Fig.~10,
clearly reveals the interplay between the surface order-disorder
phase transition and the percolating properties of the adsorbed
phase at $2/3$ monolayer coverage. Namely, $i)$ the ZZ ordered phase
represents the state of the adlayer at percolation threshold and $K
\rightarrow \infty$ and $ii)$ the curve $\theta_c$ vs. $K$ crosses
the coexistence line on the temperature-concentration phase diagram
at ($\theta_c=2/3, k_B T_c /w=0.181$) and penetrates in the
ZZ-region. The last point represents the main difference between
Model IV and the other models previously analyzed. A systematic
analysis of critical exponents was not carried out since this was
out of the scope of the present work.

Since properties of adsorbed layers are often determined by
measuring adatom coverage versus adsorbate gas pressure, it is
important to show what adsorption isotherms would look in this case.
Thus, Fig.~12 shows a set of adsorption isotherms for dimers and
different values of repulsive nearest-neighbor interactions together
with the percolation line. Since the symmetry particle-vacancy,
valid for monoatomic species, is broken for dimers, the adsorption
isotherms are not symmetric around $\theta =0.5$. In addition, two
well defined and pronounced steps appears as $K$ increases. At
$\theta=0.5$, a well defined array of dimers resembling a $c(2
\times 2)$ phase, is found. The ordered structure is characterized
by a repetition of alternating files of adsorbed dimers separated by
$2$ adjacent empty sites. As the chemical potential $\mu$ increases
and $ \theta$ approaches $\theta = 2/3$, incoming dimers are
adsorbed forming domains of parallel zig-zag rows (ZZ phase) as it
was discussed above. These structures are clearly evidence of a low
temperature ordered phase. In fact, the systems undergoes continuous
phase transitions, from disorder to ordered structures
\cite{PRB4,PRB5,EURLET}.

In summary, we presented a model to investigate the process of
adsorption of interacting monomers on square, honeycomb and
triangular lattices and studied the percolating properties of the
adsorbed phase. By using Monte Carlo simulation and finite-size
scaling theory, we obtained the percolation thresholds for different
values of concentration and temperature. From this analysis, a
critical curve in the $\theta-T$ space was addressed. The line
separating the percolating and non-percolating regions was explained
in terms of simple considerations related to the interactions
present in the problem.

\section*{Acknowledgements}

This work was supported in part by CONICET (Argentina) and the
Universidad Nacional de San Luis (Argentina) under projects PIP 6294
and 322000, respectively. The numerical work were done using the
BACO parallel cluster (composed by  60 PCs each with a 3.0 MHz
Pentium-4 processors) located  at Laboratorio de Ciencias de
Superficies y Medios Porosos, Universidad Nacional de San Luis, San
Luis, Argentina.

\newpage
\section*{Figure Captions}
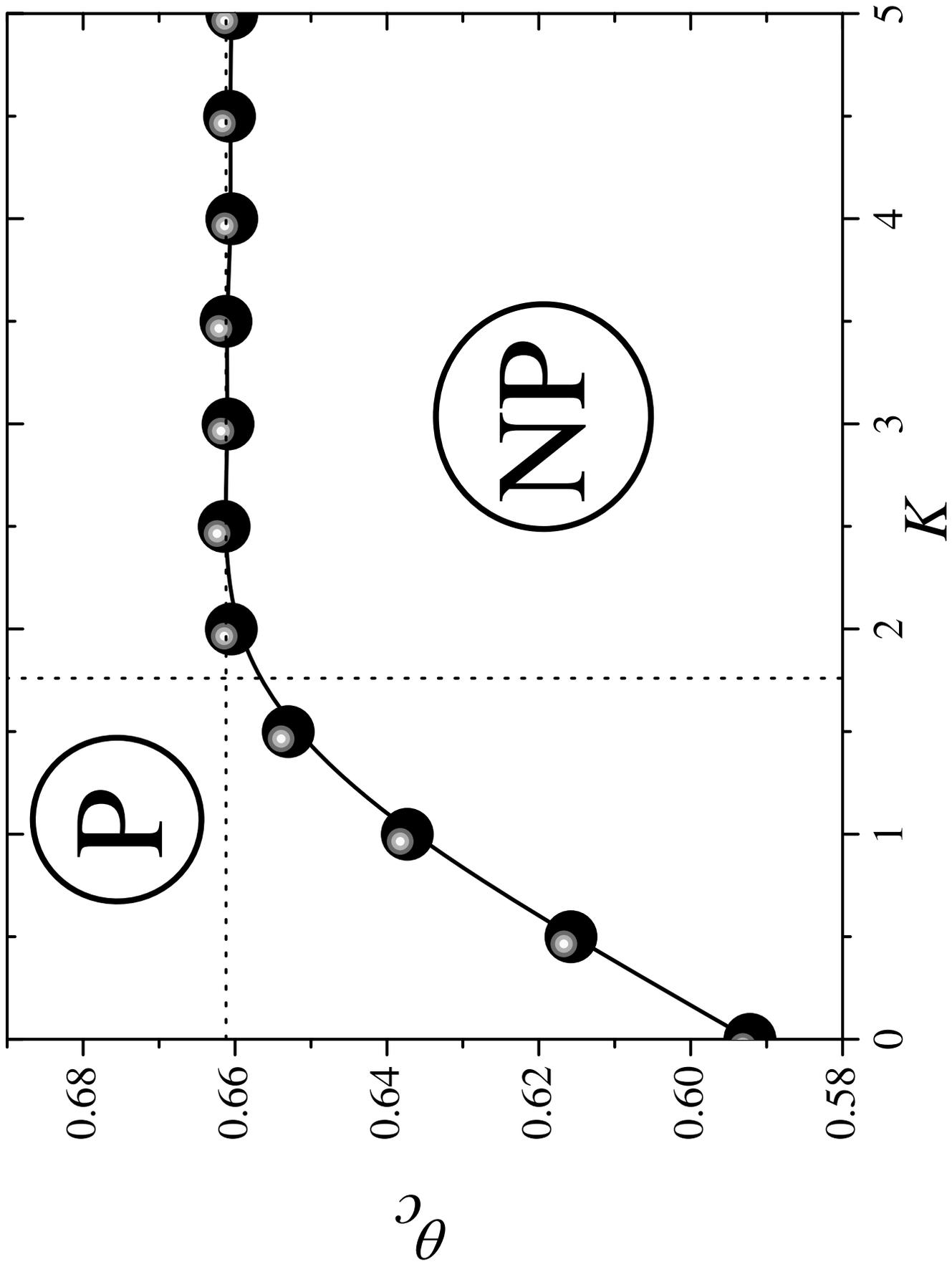
\noindent Fig. 1: Phase diagram, $\theta_c$ versus $K$, which shows
the curve separating the percolating and nonpercolating regions for
the case of monomers on a square lattice. The vertical dashed line
at $K = 1.76$ denotes the reduced critical temperature for the phase
transition occurring in the adlayer phase for repulsive interacting
particles. Horizontal dashed line at $\theta_c = 0.662$ is the
critical coverage at saturation regime for $K > 1.76$. The error
bars are smaller than the symbol size.

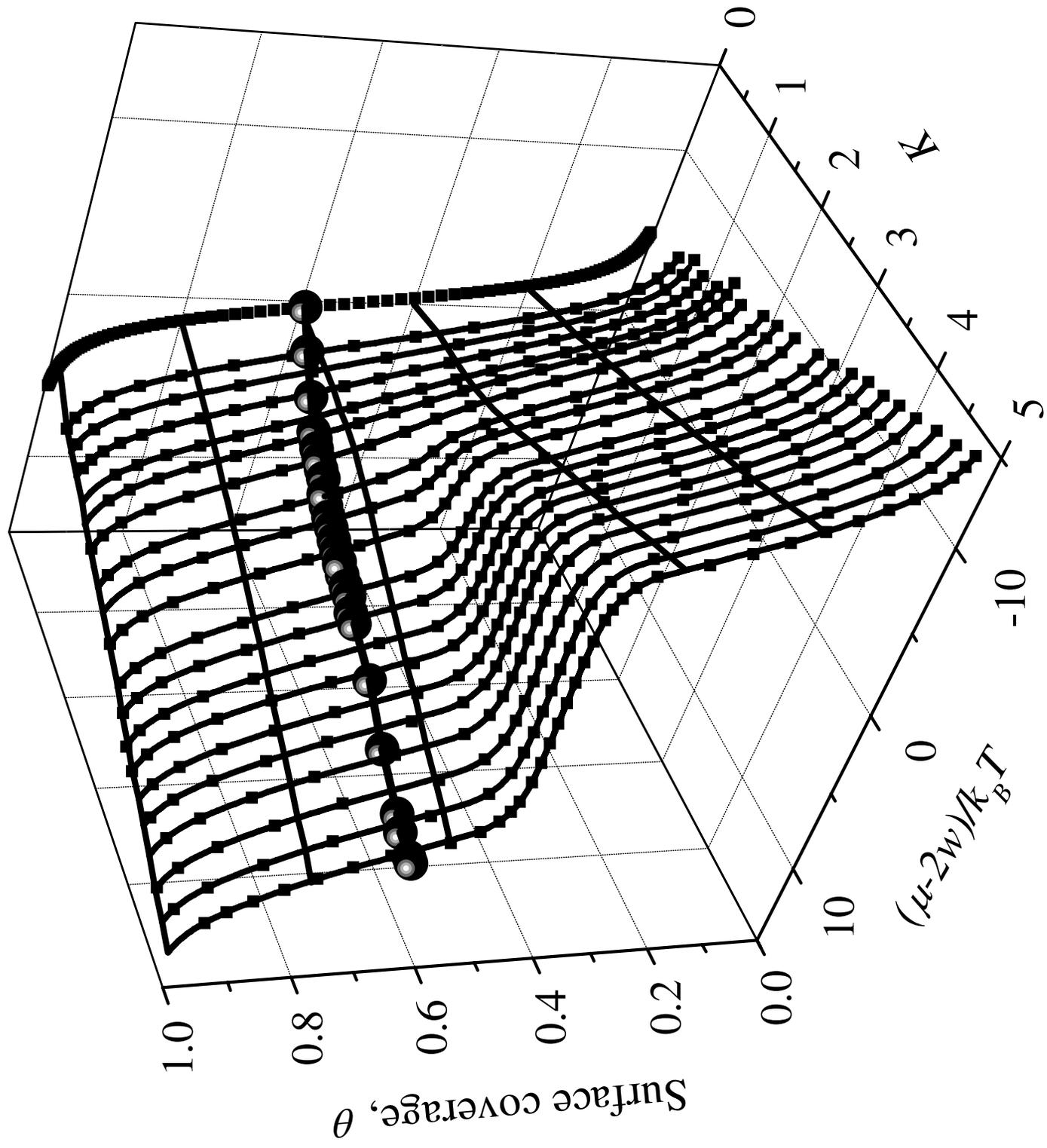
\noindent Fig. 2: Adsorption isotherms for Model I (surface
coverage, $\theta$, vs. normalized chemical potential, $\mu /k_BT$
and reciprocal temperature expressed in units of $K$). Solid lines
are obtained by the RSRG method, small symbols denote MC data while
large spheres are the same points as in Fig.~1.

\noindent Fig. 3: Phase diagram (critical temperature versus surface
coverage) corresponding to Model I obtained by RSRG, solid line.
Phase diagram, $K^{-1}$ versus $\theta_c$, which shows the curve
separating the percolating and nonpercolating regions, spheres. The
inset is a snapshot of the ordered phase c(2x2).

\noindent Fig. 4: Phase diagram, $\theta_c$ versus $K$, for Model
II.  The critical coverage at saturation regime is $\theta_c =
0.758$. The error bars are smaller than the symbol size.

\noindent Fig. 5: Adsorption isotherms for Model II and different
values of $K$, as indicated with small symbols. Solid lines are
obtained by the RSRG method, small symbols denote MC data while
large spheres are the same points as in Fig.~4.

\noindent Fig. 6: Phase diagram (critical temperature versus surface
coverage) corresponding to Model II obtained by RSRG, solid line.
Phase diagram, $K^{-1}$ versus $\theta_c$, which shows the curve
separating the percolating and nonpercolating regions, spheres. The
inset is a snapshot of the ordered phase.

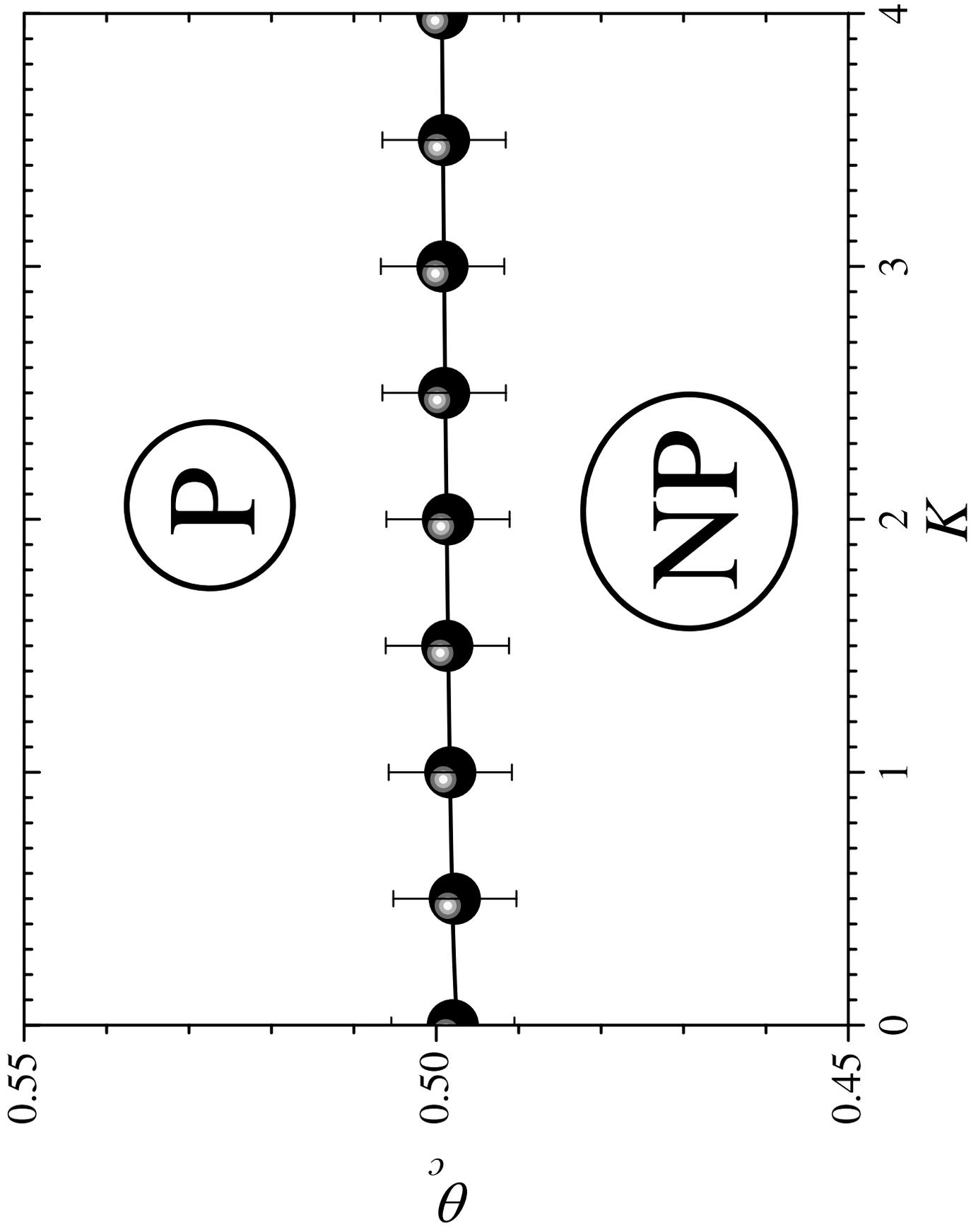
\noindent Fig. 7: Phase diagram, $\theta_c$ versus $K$, for Model
III.  The error bars are included in the figure.

\noindent Fig. 8: Phase diagram (critical temperature versus surface
coverage) corresponding to Model III obtained by RSRG, solid line.
It consists of two symmetrical curves around $\theta=0.5$. The first
one, for $\theta < 0.5$ limits the ordered phase where the particles
are arranged in such a way that pairs of particles on
nearest-neighbor lattice sites are not present (which is equivalent
to an ordered $\uparrow\downarrow\downarrow $ phase using magnetic
language). The second one, reflects the ordered phase where two of
the sublattices are occupied ($\uparrow\uparrow\downarrow $). The
insets are snapshots of the corresponding ordered phases. Phase
diagram, $K^{-1}$ versus $\theta_c$, which shows the curve
separating the percolating and nonpercolating regions, spheres.

\noindent Fig. 9: Adsorption isotherms for Model III and different
values of $K$, as indicated with small symbols. Solid lines are
obtained by the RSRG method, small symbols denote MC data while
large spheres are the same points as in Fig.~7.

\noindent Fig. 10: Phase diagram $\theta_c$ vs. $K$ for Model IV
(spheres) plotted together with data of Model I for comparison (open
circles).

\noindent Fig. 11: Phase diagram (critical temperature versus
surface coverage) corresponding to Model IV, solid line. It consists
of curves representing the regions where the phases plotted in the
insets are stables. Phase diagram, $K^{-1}$ versus $\theta_c$, which
shows the curve separating the percolating and nonpercolating
regions, spheres.

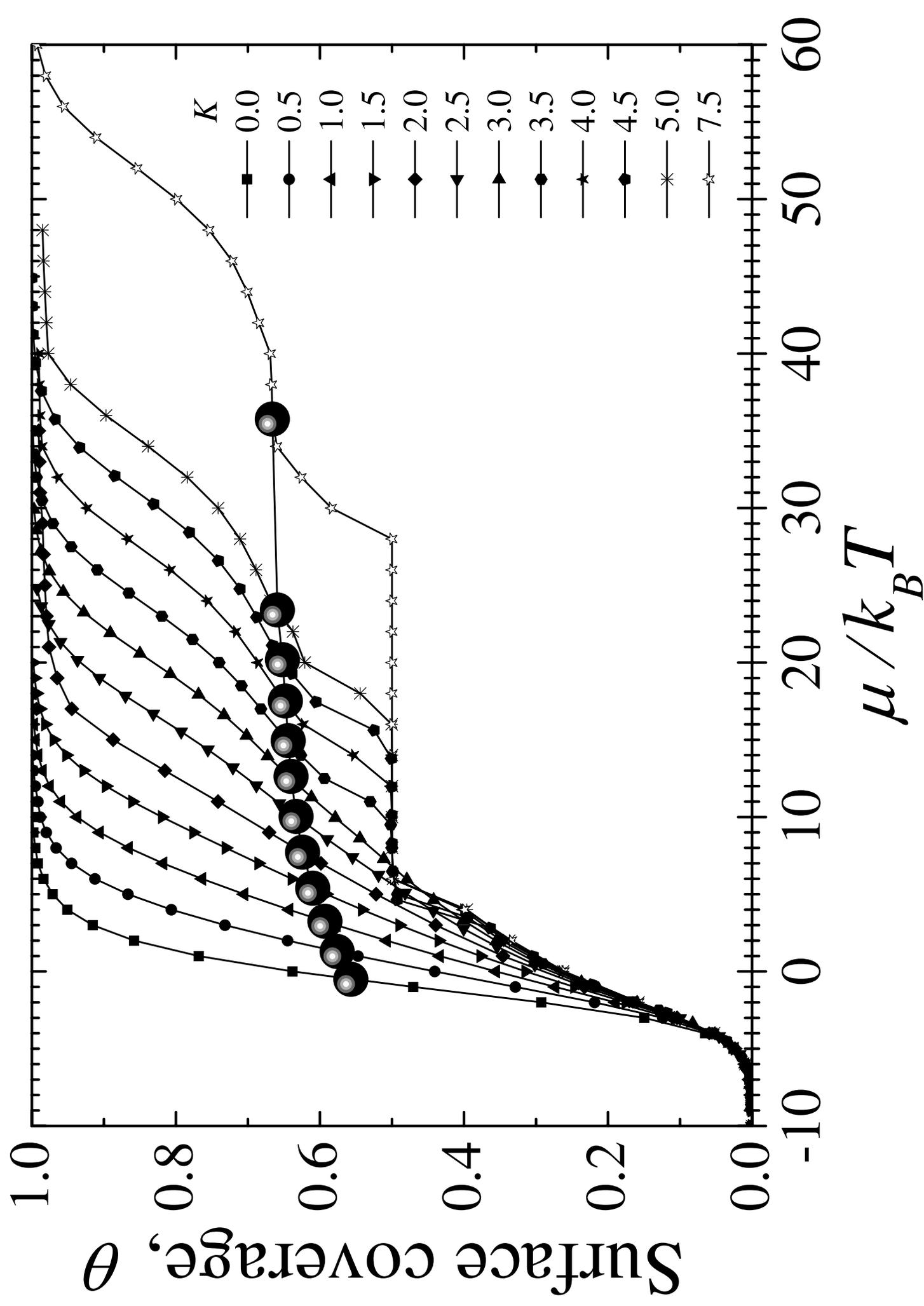
\noindent Fig. 12: Adsorption isotherms for Model IV and different
values of $K$, as indicated. Small symbols denote MC data while
large spheres are the same points as in Fig.~10.


\end{large}

\newpage
\begin{thebibliography}{9}
\parskip 0cm
\bibitem{coniglio1} A. Coniglio, J. Phys.: Condens. Matter {\bf 13}, 9039 (2001).
\bibitem{coniglio2} A. Coniglio, Nuclear Physics A {\bf 681}, 451
(2001).
\bibitem{coniglio3} A. Coniglio, Physica A {\bf 281}, 129 (2000).
\bibitem{FK} C. M. Fortuin and P. W. Kasteleyn, Physica {\bf 57}, 536 (1972)
\bibitem{FK01} P. W. Kasteleyn and C. M. Fortuin, J. Phys. Soc. Japan Suppl. {\bf 16}, 11
(1969).
\bibitem{Wu01} F. Y. Wu, Rev. Mod. Phys. {\bf 54}, 235 (1982).
\bibitem{Hill} T.L. Hill, {\it Statistical Mechanics} (New York, Mc Graw-Hll, 1956).
\bibitem{Fisher} M.E. Fisher, Physics (N.Y.) {\bf 3}, 225 (1967) .
\bibitem{fortu1} S. Fortunato, Phys. Rev. B {\bf 66}, 054107 (2002).
\bibitem{fortu2} S. Fortunato, Phys. Rev. B {\bf 67}, 014102 (2003).
\bibitem{gimenez} M. C. Gim\'enez, F. Nieto, A.J. Ramirez-Pastor, J. Phys. A: Math. Gen. {\bf 38}, (2005) 3253.
\bibitem{RUDEVE} W. Rudzi\'nski and D. H. Everett,
{\it Adsorption of Gases on Heterogeneous Surfaces} (Academic Press,
New York, 1992).
\bibitem{SURFSCI3} A. J. Ramirez-Pastor,  J. L. Riccardo and V. Pereyra, Surf. Sci. {\bf 411}, 294 (1998).
\bibitem{PRB3} A. J. Ramirez-Pastor, T. P. Eggarter, V. D. Pereyra and J. L. Riccardo, Phys. Rev. B {\bf 59}, 11027 (1999).
\bibitem{PRB4} F. Rom\'a, A. J. Ramirez-Pastor and J. L. Riccardo, Phys. Rev. B {\bf 68}, 205407 (2003).
\bibitem{PRB5} F. Rom\'a, A. J. Ramirez-Pastor and J. L. Riccardo, Phys. Rev. B {\bf 72}, 035444 (2005).
\bibitem{kawasaki} K. Kawasaki, in C. Domb and M. Green, editors, {\it Phase Transitions and Critical Phenomena},
 Vol. 2, (Academic, London, 1972).
\bibitem{Stauffer} D. Stauffer, {\it Introduction to Percolation Theory}, (Taylor \& Francis, London, 1985).
\bibitem{Sahimi} M. Sahimi, {\it Application of the percolation theory}, (Taylor \& Francis, London, 1992).
\bibitem{Zallen} R.Zallen, {\it The Physics of Amorphous Solids}, (John Willey \& Sons, NY, 1983).
\bibitem{Essam} J.W. Essam, Report on Progress in Physics {\bf 43}, 843 (1980).
\bibitem{Hovi} J.-P. Hovi and A. Aharony, Phys. Rev. B {\bf 53}, 235 (1996).
\bibitem{Binder} K. Binder, Reports on Progress in Physics {\bf 60}, 488 (1997).
\bibitem{YONE1} F. Yonezawa, S. Sakamoto and M. Hori, Phys. Rev. B {\bf 40}, 636, (1989).
\bibitem{YONE2} F. Yonezawa, S. Sakamoto, and M. Hori, Phys. Rev. B {\bf 40}, 650, (1989).
\bibitem{COR1} V. Cornette, A.J. Ramirez-Pastor and F. Nieto, Physica A {\bf 327}, 71 (2003).
\bibitem{COR2} V. Cornette, A.J. Ramirez-Pastor and F. Nieto, Eur. Phys. J. B {\bf 36}, 391 (2003).
\bibitem{DOLZ} M. Dolz, F. Nieto, A.J. Ramirez-Pastor, Eur. Phys. J. B {\bf 43}, 363 (2005).
\bibitem{DOLZ01} M. Dolz, F. Nieto, A.J. Ramirez-Pastor, Phys. Rev. E {\bf 72}, 066129 (2005).
\bibitem{Watts} G.M.T. Watts, J. Phys. A: Math. Gen. {\bf 29}, L363 (1996).
\bibitem{Lag01} R. Langlands, P. Pouliot and Y. Saint-Aubin, Bull. Am. Math. Soc. {\bf
30}, 1 (1994).
\bibitem{Lag02} R. Langlands, C. Pichet, P. Pouliot and Y.
Saint-Aubin, J. Stat. Phys. {\bf 67}, 553 (1992).
\bibitem{Cardy} J.L. Cardy, J. Phys. A: Math. Gen. {\bf 25}, L201 (1992); Nucl. Phys. B {\bf 324}, 581 (1989);
 Nucl. Phys. B {\bf 275}, 200 (1986).
\bibitem{Nicholson} D. Nicholson and N. G. Parsonage, {\it Computer Simulation
and Statistical Mechanics of Adsorption}, (Academic Press, London,
1982).
\bibitem{Metropolis} N. Metropolis, A. W. Rosenbluth, M. N. Rosenbluth, A. W. Teller,
E. Teller, J. Chem. Phys. {\bf 21}, 1087 (1953).
\bibitem{Wilson} K.G. Wilson , Rev. Mod. Phys. {\bf 47} 773 (1975).
\bibitem{Nie75}  Th. Niemeijer and J. M. J. van Leeuwen, {\it %
Renormalization Theory for Ising-like Spin Systems }in {\it \ Phase
Transitions and Critical Phenomena}, edited by C. Domb and M. S.
Green (Academic, New York, 1976), Vol.6, Chap.7.
\bibitem{Mahan} G. D. Mahan and F. H. Claro, Phys.Rev. B {\bf 16}, 1168
(1977).
\bibitem{BerkerI} A.N. Berker, S. Ostlund, and F.A. Putnam, Phys. Rev. B {\bf 17},
3650 (1978).
\bibitem{BerkerII}   R.G. Caflisch and A.N. Berker, Phys. Rev. B {\bf 29}, 1279 (1984).
\bibitem{BerkerIII} R.G. Caflisch, A.N. Berker, and M. Kardar, Phys. Rev. B {\bf 31}, 4527
(1985).
\bibitem{Nie74}  Th. Niemeyer and J. M. J. van Leeuwen, Physica {\bf 71}, 17
(1974).

\bibitem{Nau74}  M. Nauenberg and B. Nienhuis, Phys.Rev.Lett. {\bf 33}, 1598
(1974).

\bibitem{Nau75}  B. Nienhuis and M. Nauenberg, Phys.Rev.Lett. {\bf 35}, 477
(1975).
\bibitem{sl}  A.A. Tarasenko, L. Jastrabik, F. Nieto and C. Uebing,
Phys. Rev. B {\bf 59}, 8252 (1999); Physical Chemistry Chemical
Physics (PCCP) {\bf 1}, 1583 (1999); A.A. Tarasenko, F. Nieto, L.
Jastrab�k and C. Uebing. The European Physical Journal D {\bf 12},
311 (2000).
\bibitem{tl} A.A. Tarasenko, F. Nieto and C. Uebing. Physical Chemistry Chemical Physics (PCCP), {\bf 2} 3453 (2000);
A.A. Tarasenko, F. Nieto, L. Jastrabik and C. Uebing. Phys. Rev. B
{\bf 64} 0754131 (2001); Surface Science {\bf 536}, 1 (2003) 1-14;
F. Nieto, A.A. Tarasenko, "Collective surface diffusion of
interacting particles on a triangular lattice: real-space
renormalization group and Monte Carlo approaches" in "Trends in
Surface Science Research", Editor: Charles P. Norris, Nova Science
Publishers, Inc., New York, (2005).
\bibitem{hl} A. A. Tarasenko, L. Jastrabik and  C. Uebing Phys. Rev. B {\bf 57}, 10166
(1998).
\bibitem{Sch76} M. Schick, J. S. Walker and M. Wortis, Phys.Lett. A {\bf 58}, 479 (1976).
\bibitem{Sch77} M. Schick, J. S. Walker and M. Wortis, Phys. Rev. B {\bf 16}, 2205 (1977).
\bibitem{Kin81} W. Kinzel and M. Schick, Phys. Rev. B {\bf 23}, 3435 (1981).
\bibitem{Met73} B.D. Metcalf, Phys. Lett. A {\bf 45}, 1 (1973).
\bibitem{EURLET} F. Rom\'a, J. L. Riccardo and A.J. Ramirez-Pastor, {\it Critical behavior of
repulsive dimers at $2/3$ monolayer coverage}, submitted (2006).
\end{thebibliography}
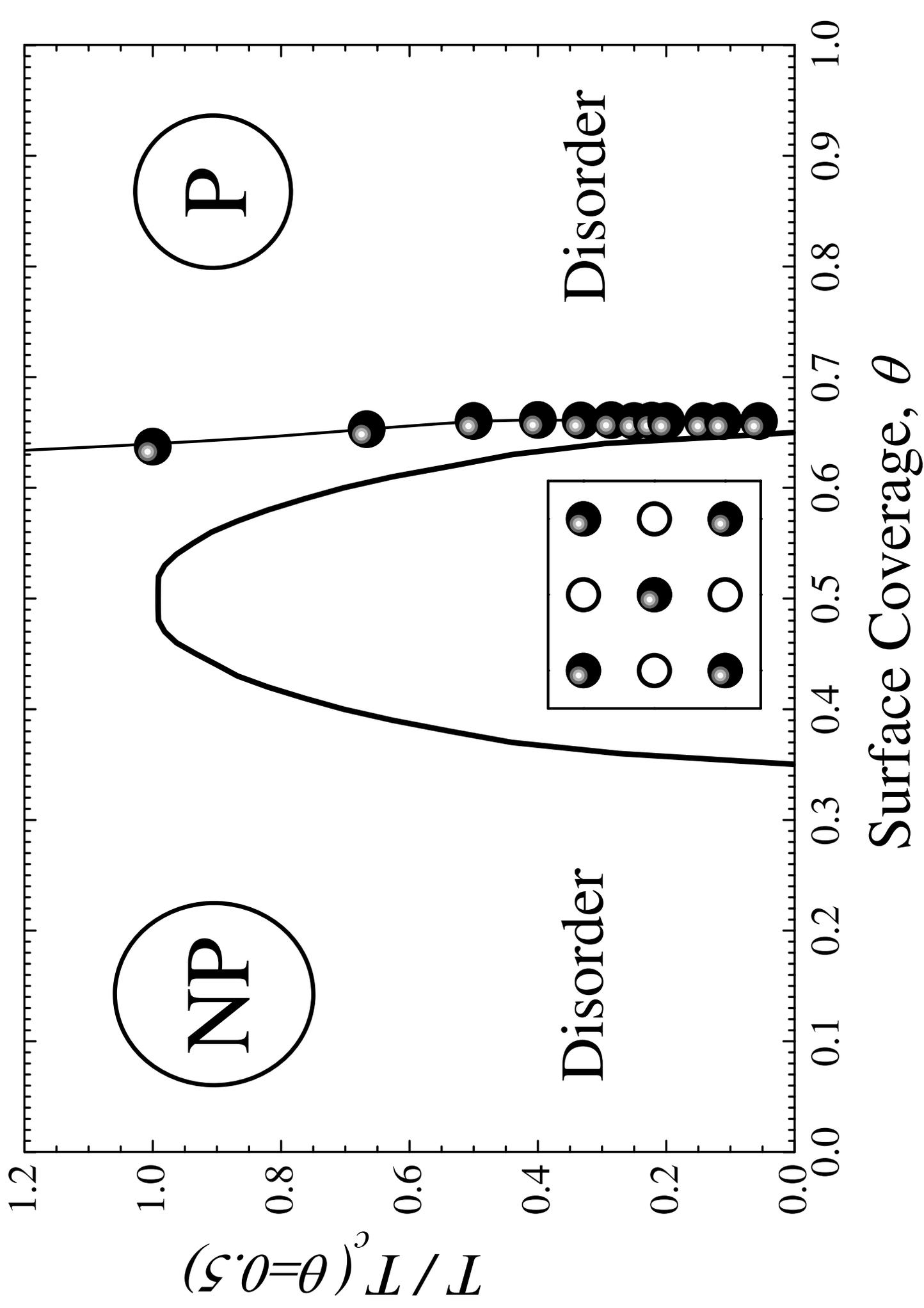
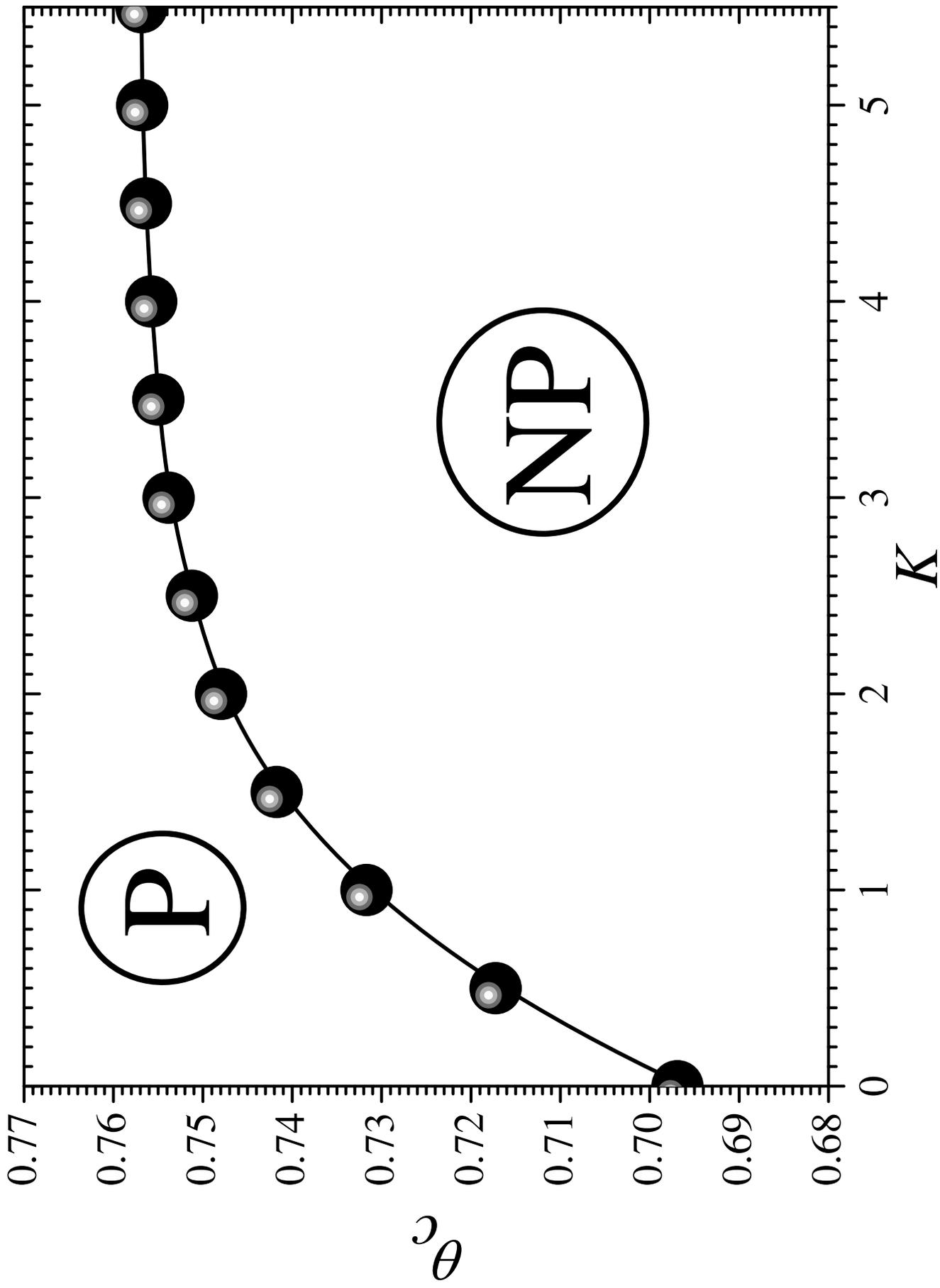
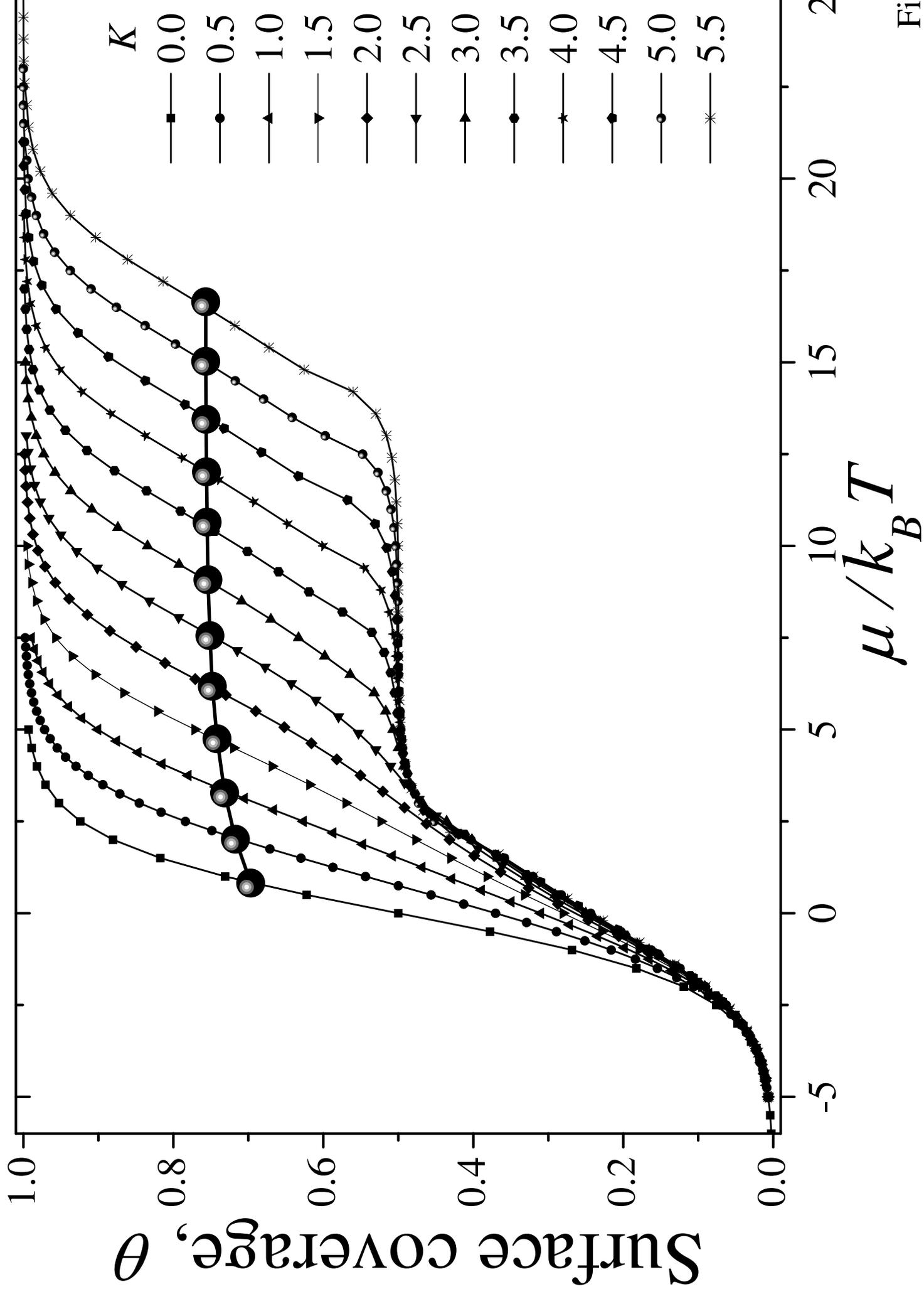
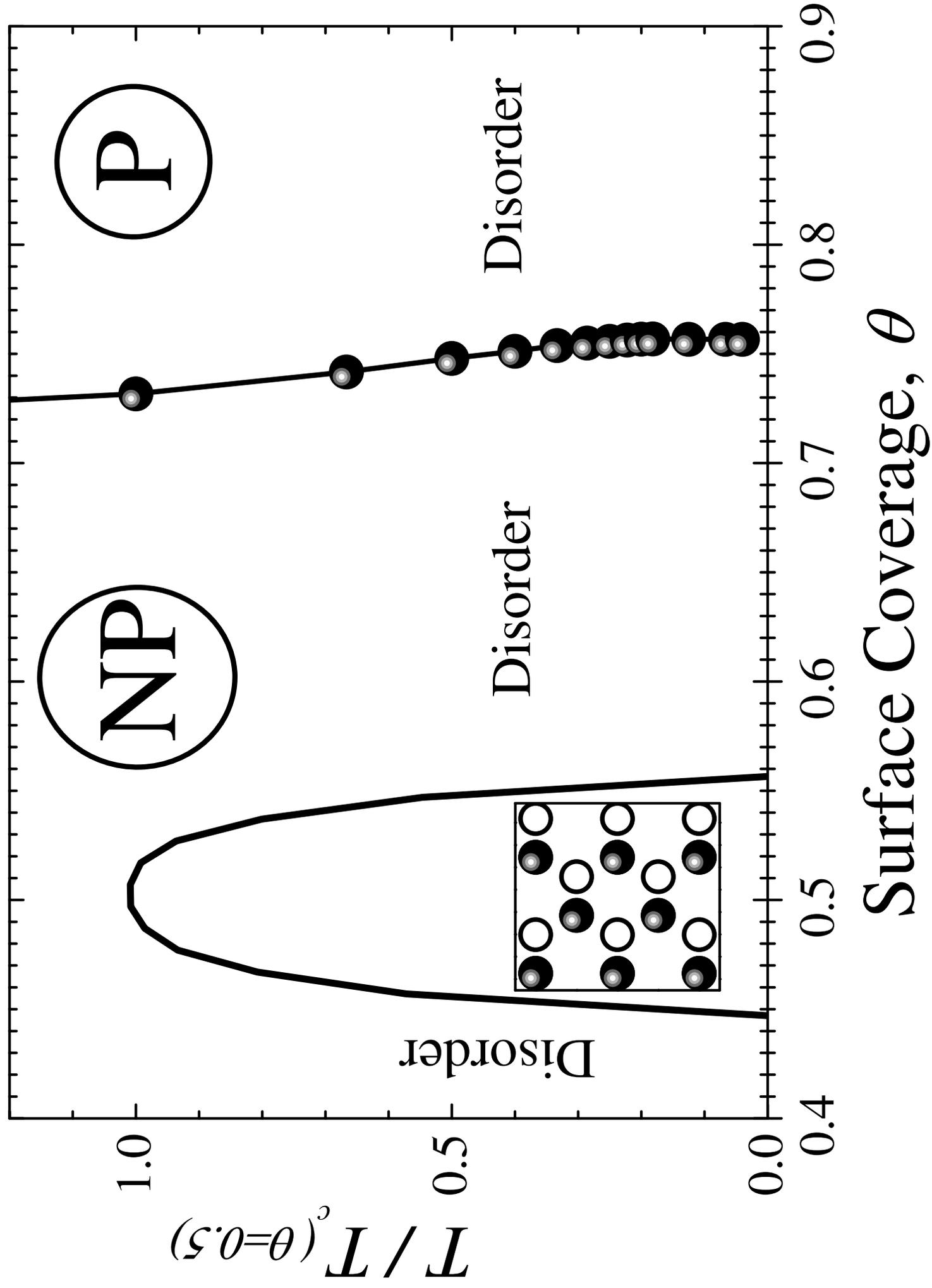
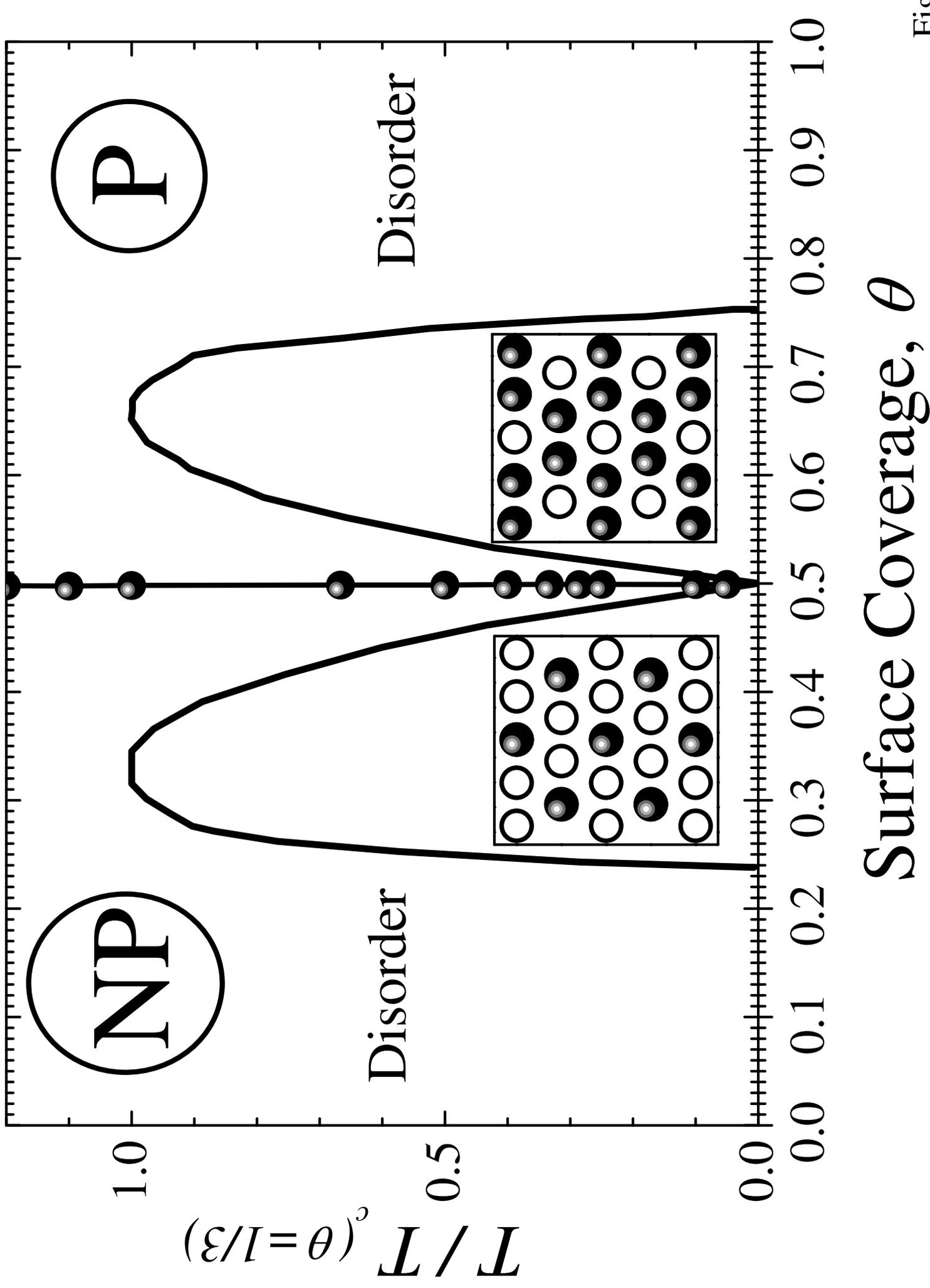
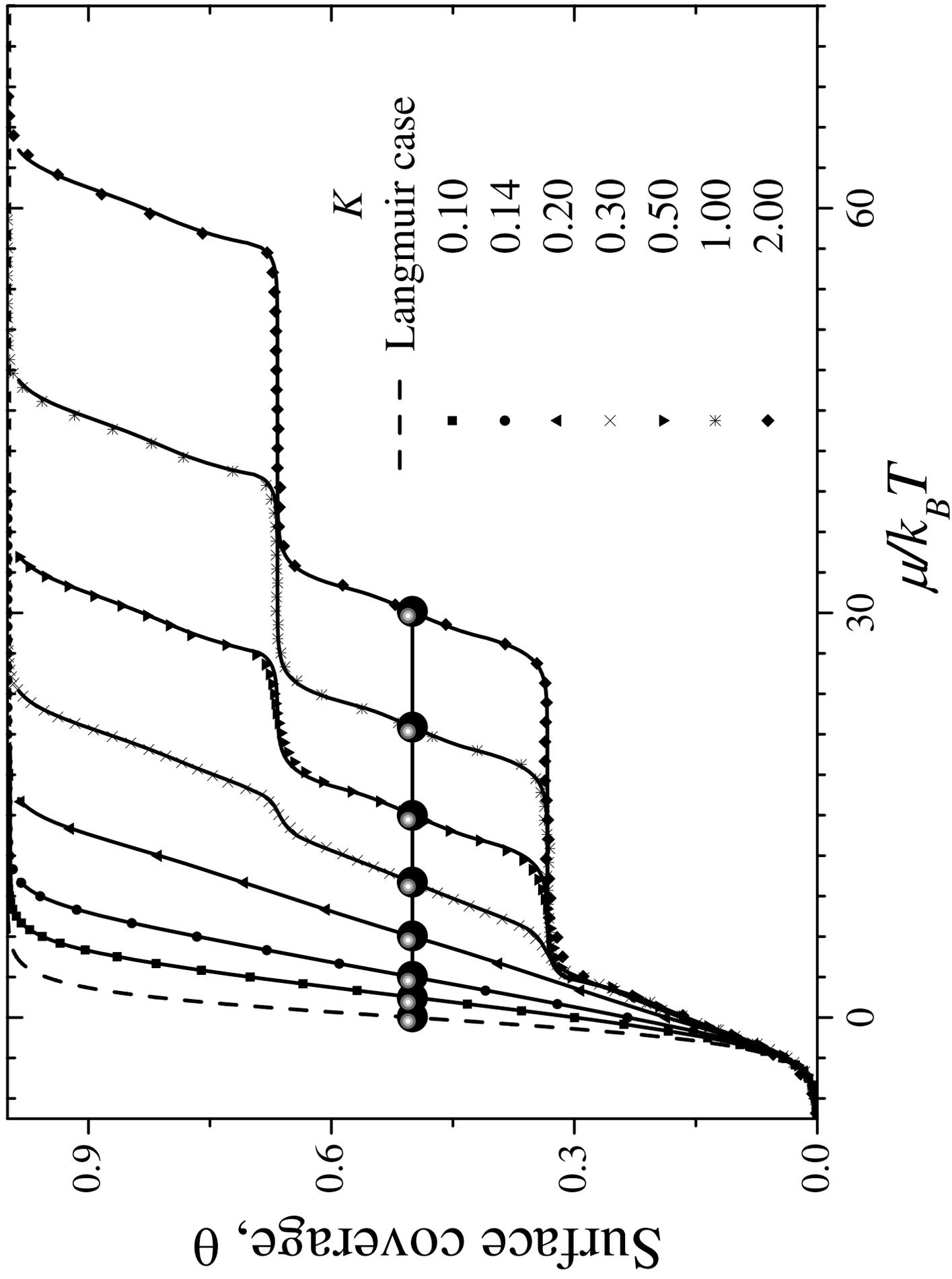
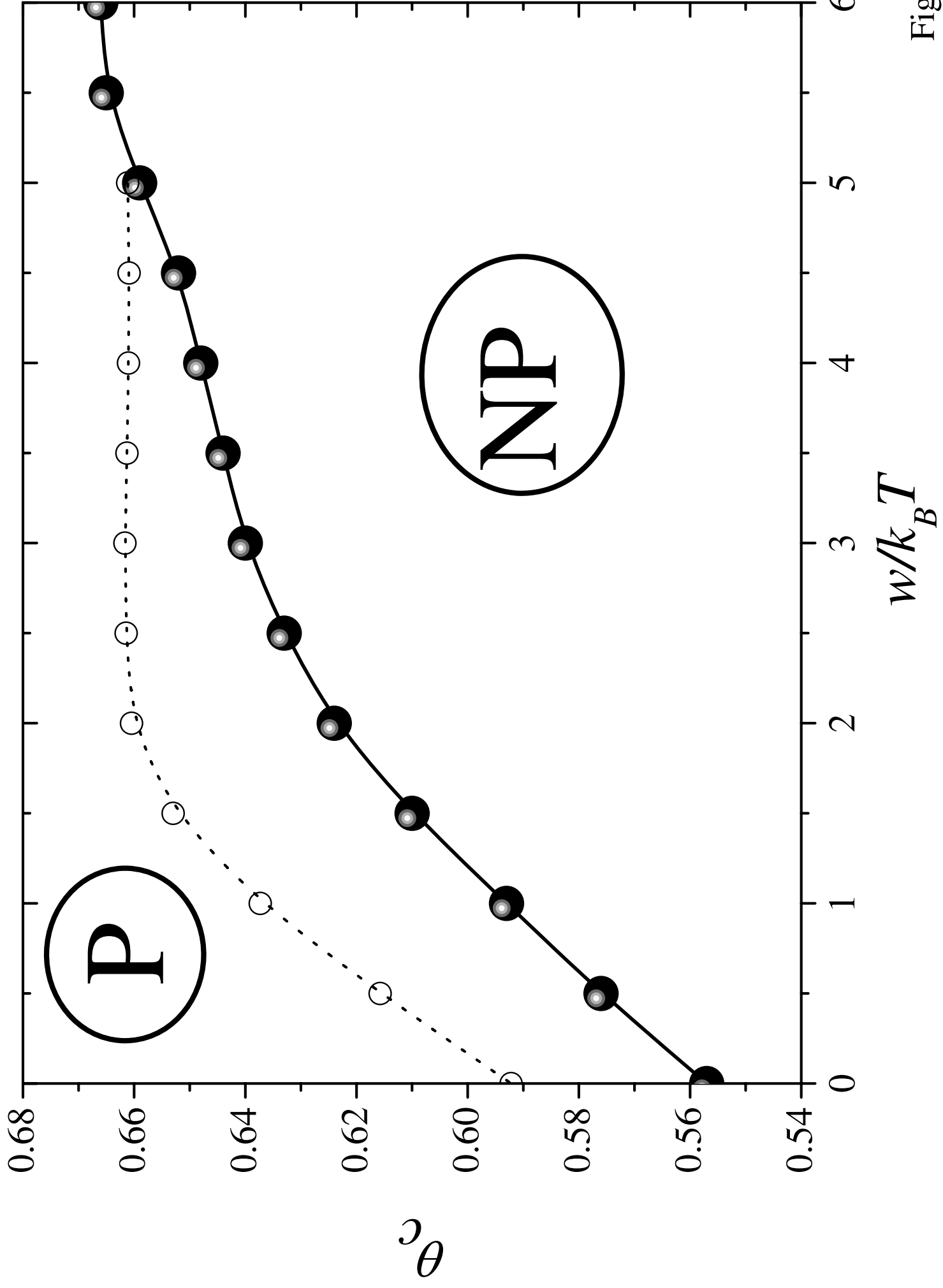
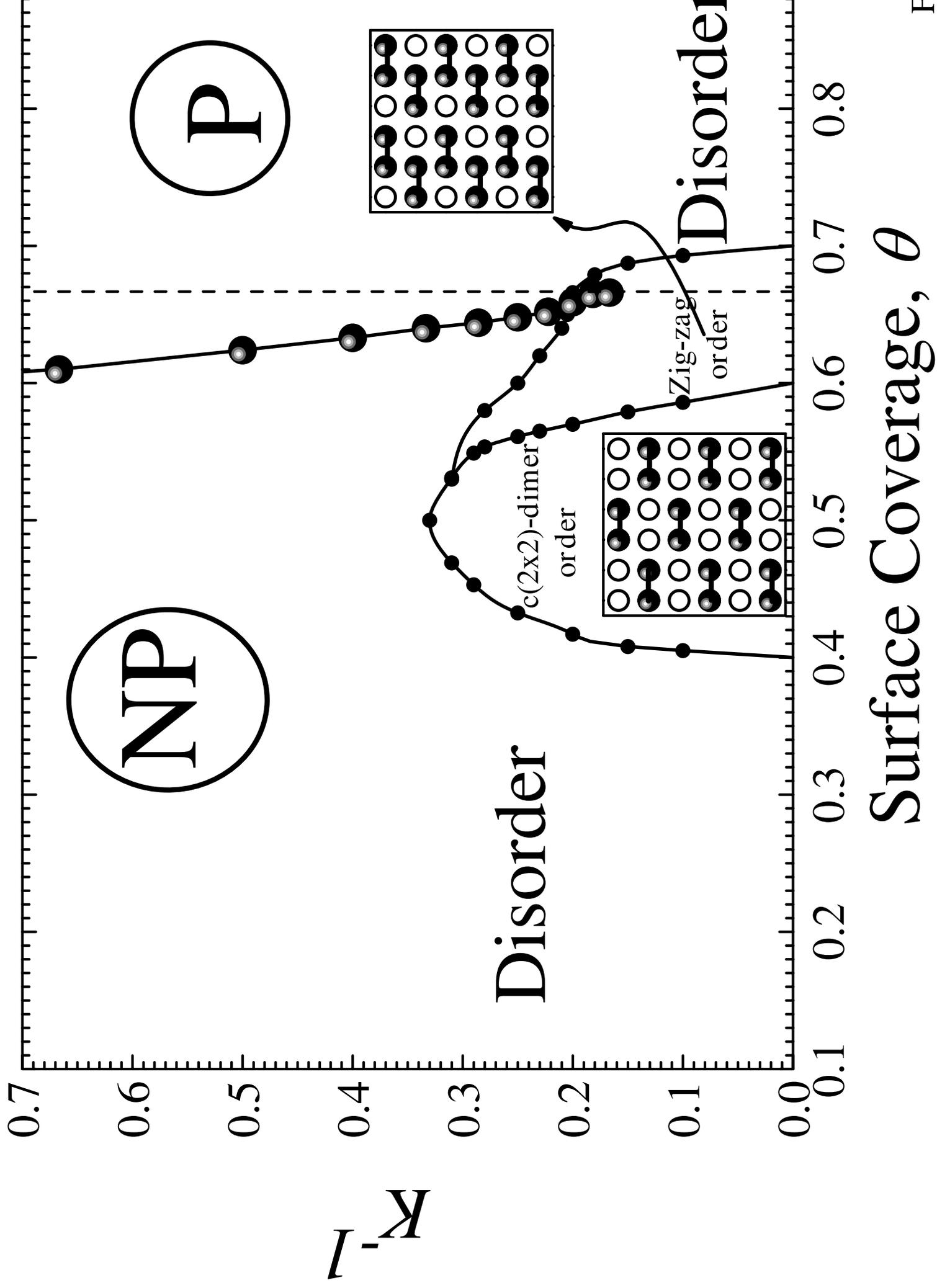
\end{document}